\documentclass[prb,showpacs,superscriptaddress,twocolumn]{revtex4}
\usepackage{epsfig}
\usepackage{graphicx}

\newcommand{\be}{\begin{equation}}
\newcommand{\ee}{  \end{equation}}
\newcommand{\ba}{\begin{eqnarray}}
\newcommand{\ea}{  \end{eqnarray}}
\newcommand{\bi}{\begin{itemize}}
\newcommand{\ei}{  \end{itemize}}

\begin{document}

\title{Nonlinear electronic transport in nanoscopic devices: \\
Nonequilibrium Green's functions versus scattering approach}

\author{A. R. Hern\'andez}
\affiliation{Laborat\'orio Nacional de Luz S\'{\i}ncrotron,
             Caixa Postal 6192, 13083-970 Campinas, Brazil}
\affiliation{Centro Brasileiro de Pesquisas F\'{\i}sicas,
             R.~Dr.~Xavier Sigaud 150, 22290-180 Rio de Janeiro, Brazil}

\author{C. H. Lewenkopf}
\affiliation{Departamento de F\'{\i}sica Te\'orica, Universidade do Estado do Rio de Janeiro, 20550-900 Rio de Janeiro, Brazil}

\date{\today}

\begin{abstract}
We study the nonlinear elastic quantum electronic transport properties of nanoscopic devices using the Nonequilibrium Green's function (NEGF) method. The Green's function method allows us to expand the $I-V$ characteristics of a given device to arbitrary powers of the applied voltages. By doing so,
we are able to relate the NEGF method to the scattering approach, showing their similarities
and differences and calculate the conductance coefficients to arbitrary order.
We demonstrate that the electronic current given by NEGF is gauge invariant  to all orders in powers of $V$, and discuss the requirements for gauge invariance in the standard Density Functional Theory (DFT) implementations in molecular electronics. We also analyze the symmetries of the nonlinear conductance coefficients with respect to a magnetic field inversion and the violation of the Onsager reciprocity relations with increasing source-drain bias.
\end{abstract}


\maketitle
\section{Introduction}
\label{sec:introduction}

There is a growing experimental and theoretical interest in quantum nonlinear
electronic transport properties of nanoscopic devices. Experiments in mesoscopic
semiconductors, such as quantum dots, \cite{Lofgren04,Marlow06,Zumbuhl06} and
quantum rings, \cite{Leturcq06,Angers07} have focused on the investigation of
rectification effects and violations of the Onsager-Casimir reciprocity
relations. Similar issues have also been examined in the electronic transport
through carbon nanotubes.\cite{Wei05}
Nonlinear transport is also of major interest in molecular electronics, \cite{Nitzan03}
where considerable experimental effort has been put in showing that single molecules
can be used as diodes, transistors, and switches.

The development of a comprehensive quantum nonlinear electronic transport theory,
a non-equilibrium quantum many-electron problem, is still quite a challenge.
Notwithstandingly, significant advances have been already achieved, particularly by
restricting the theoretical analysis to elastic processes. Within this approximation,
theoretical progress has mainly been achieved by pursuing two apparently very different
paths, namely, the scattering approach put forward by B\"uttiker and collaborators
\cite{Buttiker93,Christen96,Sanchez04,Polianski06,Polianski07} and the nonequilibrium
Green's function method. \cite{Wang99,Xue02,Spivak04,Deyo06}

In both approaches the current-voltage $I-V$ characteristics is written in terms of
transmission coefficients that account for the potential landscape built in the devices
due to the applied bias. The scattering approach \cite{Buttiker93,Christen96} casts
the current as a power series of the bias. The $S$-matrix serves not only to
compute the transmission, like in the Landauer formula, but also to calculate the
electrostatic potential built in the conductor by means of physical considerations.
Those guarantee that the electrostatic potential is gauge invariant order by order
in powers of $V$. Alternatively, NEGF has also been used to investigate the linear
and non-linear transport properties of mesoscopic \cite{Wang99,Spivak04,Deyo06} and
molecular systems. \cite{Demle01,Taylor01,Xue02,Evers03,Ke04}
Here, the current and the electrostatic potential are calculated self-consistently to
all orders at once. The formalism is quite powerful and robust but, as it is often
the case in self-consistent calculations, different physical processes become
inextricable making difficult to understand their role and importance for the
electronic transport.

These considerations raise a natural question: To what extent are these approaches
similar? One of the main purposes of this paper is to answer this question and explicitly
show that both approaches are, in principle, equivalent. Furthermore, we show that
differences appear depending on how the underlying many-body electronic problem is
approximated.

To this end we use NEGF to write the current flowing through a multi-lead
elastic conductor as a power series of the bias $V$, as done in Ref.~\onlinecite{Wang99}.
Treating the many-electron problem in the Hartree approximation we explicitly show that,
in the Thomas-Fermi limit, the leading nonlinear correction in the $I-V$ characteristics
reduces, almost exactly, to the scattering approach result. We discuss gauge invariance
and the Onsager-Casimir reciprocity relations. At every step, we also analyze these
symmetries beyond the Hartree term by addressing the standard Density Functional
Theory (DFT) implementation for molecular electronics.\cite{Xue02} We stress that, in
contrast to the scattering approach, the NEGF formalism is not restricted to a local
approximation. The many-body nature of NEGF allows one to extend its standard
implementation in a variety of ways. To show this, we address the case where the
many-body problem is treated in the Hartree-Fock approximation. \cite{GolGefen}

The presentation of the paper is organized as follows:
In Sec.~\ref{sec:scattering} we present the scattering approach highlighting its main
elements and results for later comparison with the NEGF approach.
In Sec.~\ref{sec:model} we present the model Hamiltonian considered in this study. In
Sec. \ref{sec:I-V} we calculate the transmission coefficients and the electrostatic
potential using nonequilibrium Green's functions. By these means, it is possible to
systematically calculate the conductance coefficients and the characteristic potentials
self-consistently to all orders in $V$, as discussed in Sec.~\ref{sec:expansion}.
The similarities between both approaches are discussed in Sec.~\ref{sec:connectionButtiker},
where we also show how differences appear. The NEGF implementation of the Hartree-Fock
approximation is described in Sec.~\ref{sec:HF}. Finally, our conclusions are presented
in Sec.~\ref{sec:conclusions}.

\section{The scattering approach}
\label{sec:scattering}

Let us consider a conductor connected by leads to $\alpha =1,\cdots, N$ electronic reservoirs at a temperature $T$. In the absence of an applied bias,
the system is in thermal and chemical equilibrium, characterized by a chemical
potential $\mu_0$.
By applying voltages $\{V_\alpha\}$ to the reservoirs, the system is driven out
of equilibrium and an electronic current flows.

According to B\"uttiker, \cite{Buttiker93} in the absence of inelastic processes,
the current at the lead $\alpha$ is given by
\be
\label{eq:LBscattering}
I_{\alpha}=\frac{2 e}{h}\sum_{\beta=1}^N \int_{-\infty}^{\infty} \!
dE\, f_\beta(E)\, A_{\alpha\beta}(E,U({\bf r})) \;.
\ee
Here $f_\beta(E) = f_0(E-eV_{\beta})$, where $f_0(E)=(e^{E/k_{\rm B}T} + 1)^{-1}$ is
the Fermi distribution function and $k_{\rm B}$ is the Boltzmann constant.
For notational convenience, we consider $V_\beta$ as measured with respect to the
equilibrium potential $\mu_0$, namely, $V_\beta \rightarrow V_\beta - \mu_0/e$.

The transmission is identified with
\ba
A_{\alpha\beta}(E,U({\bf r}))
&=& {\rm Tr} [{\bf 1}_{\alpha}\delta_{\alpha\beta} - {\bf S}^\dagger_{\alpha\beta}{\bf S}^{}_{\alpha\beta}]\;,
\ea
where ${\bf S}_{\alpha\beta}(E,U({\bf r}))$ denotes the scattering matrix with
lines and rows associated with the transversal modes at the contact $\alpha$ and
$\beta$, respectively. ${\bf 1}_\alpha$ is the identity matrix whose rank is given
by the number of propagating channels in the contact $\alpha$. The trace runs over
all open channels in $\alpha$ and $\beta$.

The transmission coefficient $A_{\alpha\beta}$ and the scattering matrix ${\bf S}_{\alpha\beta}$
are functions of the electron energy and functionals of the electrostatic potential
$U({\bf r})$ in the conductor. In linear response, $A_{\alpha\beta}$ is computed at
the equilibrium potential $U_{\rm eq}({\bf r})$ that is established when all reservoirs
have the same chemical potential $\mu_0$. Beyond this regime, it is necessary to compute
$U({\bf r})$ self-consistently, as pointed out by Landauer.\cite{Landauer87}

To make analytical progress, it is convenient to expand all quantities in powers of $V$.
The local electrostatic potential $U({\bf r})$ reads
\ba
\label{eq3}
U({\bf r})&=& U_{\rm eq}({\bf r}) + \sum_{\alpha}u_{\alpha}({\bf r})V_{\alpha}
\nonumber\\
&& + \frac{1}{2}\sum_{\alpha\beta}u_{\alpha\beta}({\bf r})V_{\alpha}V_\beta+ O(V^3)
\ea
where  $u_{\alpha \beta \cdots}({\bf r})$ is the characteristic potential defined by
\be
\label{eq:defu_char}
u_{\alpha \beta \cdots}({\bf r})=\left( \frac{\partial}{\partial V_{\alpha}}\frac{\partial}{\partial V_{\beta}}\cdots\right) U({\bf r}) \Big|_{\{V_\gamma\}=0}.
\ee
Here $\{V_\gamma\}=0$ is a shorthand for $V_\gamma=0$, for all $\gamma$.

Some properties of the characteristic potentials follow directly from simple physical
considerations. For instance, $u_\alpha({\bf r})$ has the following properties: \cite{Buttiker93}
(a) Changes in the electro-chemical potential of the reservoir $\alpha$ should
not affect $U({\bf r})$ inside $\beta$, hence, $u_\alpha({\bf r})=0$, when
${\bf r}$ is taken inside the reservoir $\beta\neq \alpha$.
(b) For ${\bf r}$ inside the reservoir $\alpha$, $U({\bf r}) = V_\alpha$ and,
thus, $u_\alpha({\bf r})=1$.
(c) A global change of the applied potentials, $V_\alpha \rightarrow V_\alpha
+ V_0$, makes $U({\bf r}) \rightarrow U({\bf r}) + V_0$, implying the sum rule
$\sum_\alpha u_\alpha({\bf r}) = 1$ for all ${\bf r}$.

The current $I_\alpha$, written as a power series of the applied voltages,
is cast as a function of the coefficients ${\cal G}_{\alpha\beta\cdots}$, namely,
\be
\label{eq:generalcurrent}
I_{\alpha}=\sum_{\beta}{\cal G}_{\alpha \beta}V_{\beta}+
\sum_{\beta\gamma}{\cal G}_{\alpha \beta \gamma}V_{\beta}V_{\gamma}+
\sum_{\beta\gamma\delta}{\cal G}_{\alpha \beta \gamma\delta}V_{\beta}V_{\gamma}V_\delta
+\cdots\;.
\ee
In line with the standard notation, \cite{Buttiker93} we do not write $I_\alpha$ as a Taylor
series in $\{V_\alpha\}$. In Sec.~\ref{sec:expansion} we will see how such notation
determines the symmetrization of the indices $\alpha, \beta, \cdots$ in the conductance
coefficients ${\cal G}_{\alpha\beta\cdots}$.

The coefficient ${\cal G}_{\alpha\beta}$ corresponds to the linear conductance,
as given by the Landauer formula
\be
{\cal G}_{\alpha \beta}=\frac{2 e^2}{h}\int_{-\infty}^{\infty} dE
\left(-\frac{\partial f_0}{\partial E}\right) A_{\alpha\beta}(E,U_{\rm eq}({\bf r})).
\ee
Here $A_{\alpha\beta}(E,U_{\rm eq}({\bf r}))$ is the multi-lead Landauer-B\"uttiker transmission coefficient, \cite{Buttiker93} with $S$ computed using $U_{\rm eq}({\bf r})$, as standard.

The first non-linear current correction, represented by ${\cal G}_{\alpha\beta\gamma}$ reads \cite{Buttiker93,Christen96}
\ba
\label{eq:Gabc}
{\cal G}_{\alpha \beta \gamma}&=& \frac{2 e^3}{h}\int_{-\infty}^{\infty} dE \left( -\frac{\partial f_0}{\partial E}\right)
\nonumber\\
&&\times \int\! d{\bf r} \,\Big[u_{\gamma}({\bf r})-\frac{1}{2}\delta_{\beta\gamma}\Big] \left. \frac{\delta A_{\alpha\beta}}{e\delta U({\bf r})}\right|_{\{V_{\alpha}\}=0}\;,
\ea
where the spatial integration is taken over the region where $\delta A_{\alpha\beta}/\delta U({\bf r})|_{\{V_{\alpha}\}=0}$ is non vanishing, namely, inside the conductor.

The above expression depends explicitly on the electrostatic potential via $u_\alpha({\bf r})$.
To determine $u_{\alpha}({\bf r})$, the formalism has to be supplemented by a self-consistent microscopic electronic structure calculation, or by an adequate approximation.
The latter was constructed in Ref.~\onlinecite{Buttiker93} by using the following argument:
The potential $U({\bf r})$ is related to the bias generated electronic density imbalance $\delta n({\bf r})$ in the conductor. In turn, $\delta n({\bf r})$ arises from the
charge injected by the leads and the induced charge in the conductor, in response to the injected one.

The injection properties of the sample are given by the injectivity, which reads
\ba
\label{eq:injec}
\frac{dn^{\rm s}({\bf r},\alpha )}{dE}&\!\!=&\!\!
-\frac{1}{2\pi i}\int_{-\infty}^{\infty} dE \left( -\frac{\partial f_0}{\partial E}\right)
\nonumber\\&&\!\!\!\times
\sum_{\beta}{\rm Tr} \left[ {\bf S}^{\dagger}_{\beta \alpha}\frac{\delta {\bf S}_{\beta \alpha}}
{e\delta U({\bf r})}-\frac{\delta {\bf S}^{\dagger}_{\beta\alpha}}{e\delta U({\bf r})}
{\bf S}_{\beta\alpha}\right],
\ea
evaluated at $\{V_\gamma\}=0$. The superscript ${\rm s}$ labels quantities obtained within the scattering approach. Since $dn^{\rm s}/dE$ includes spin degeneracy, our definition differs from the one of Ref.~\onlinecite{Buttiker93} by a factor 2.

To linear order in $V$, the induced charge density is given by
\be
dn_{\rm ind}({\bf r}) = e\sum_\alpha   \int d{\bf r}^\prime \,
\Pi({\bf r},{\bf r}^\prime)\, u_\alpha({\bf r}^\prime) dV_\alpha
\ee
where $\Pi ({\bf r},{\bf r}')$ is the Lindhard polarization function. \cite{Bruus-Flensberg2004} The scattering approach does
not provide a recipe to obtain the later. However, by recalling the relation between the Wigner-Smith time delay and the conductor density of states, $dn_{\rm ind}({\bf r})$ can be
readily written in the Thomas-Fermi approximation as
\be
\label{eq:ThomasFermi}
dn_{\rm ind}({\bf r}) = e\sum_\alpha  \frac{dn^{\rm s}({\bf r} )}{dE}\, u_\alpha({\bf r}) dV_\alpha .
\ee
The local density of states $dn^{\rm s}/dE$ is
\be
\frac{dn^{\rm s}({\bf r} )}{dE}=
\sum_{\beta} \frac{dn^{\rm s}(\beta, {\bf r} )}{dE},
\ee
where $dn^{\rm s}(\beta, {\bf r} )/dE$ is called emissivity and is given by
\ba
\!\!\!\!\!\!
\frac{dn^{\rm s}(\beta, {\bf r} )}{dE}&\!\!=&\!\!
-\frac{1}{2\pi i}\int dE \left( -\frac{\partial f_0}{\partial E}\right)
\nonumber\\&&\!\!\!\times
\sum_{\alpha} {\rm Tr}\left[ {\bf S}^{\dagger}_{\beta \alpha}\frac{\delta {\bf S}_{\beta \alpha}}{e\delta U({\bf r})}-\frac{\delta {\bf S}^{\dagger}_{\beta\alpha}}{e\delta U({\bf r})}{\bf S}_{\beta\alpha}\right].
\ea
These elements render the Poisson equation
\be
\label{eq:scatPoisson}
-\nabla^2 u_{\alpha}({\bf r})+4\pi e^2 \frac{dn^{\rm s}({\bf r} )}{dE}u_{\alpha}({\bf r}')=4\pi e^2  \frac{dn^{\rm s}({\bf r},\alpha )}{dE},
\ee
where both the density of states and the injectivity depend only on the scattering matrix.

Higher order conductance coefficients ${\cal G}_{\alpha\beta\gamma\cdots}$ can also be calculated
in a straightforward way. Obtaining self-consistent equations for the characteristic potentials
$u_{\alpha\beta\gamma\cdots}$ becomes increasingly more involved, but is still possible within the Thomas-Fermi approximation.\cite{Ma98}

In the next Sections we define a general model for a conductor and use the NEGF approach to
show how to systematically obtain the coefficients ${\cal G}_{\alpha\beta\gamma \cdots}$ and the characteristic potentials
$u_{\alpha\beta\gamma \cdots}$ to arbitrary order.

\section{Model Hamiltonian}
\label{sec:model}

We separate the system in two regions, namely, the leads (L) and the conductor (C) to write the Hamiltonian as
\be \label{eq:H}
{\cal H} = {\cal H}_{\rm L} + {\cal H}_{\rm C} + {\cal H}_{\rm LC}.
\ee
For definiteness, we introduce a surface ${\cal S}$ enclosing the conductor, to partition the model Hilbert
space.
The lead Hamiltonian reads
\be \label{eq:Hlead}
{\cal H}_{\rm L}=\sum_{k\alpha as} E^{}_{k\alpha a s} c^{\dagger}_{k\alpha a s}c^{}_{k\alpha a s},
\ee
where $k$ is the electron transversal wave number at the channel $a$ ($a=1, \cdots, N_\alpha$) in the lead $\alpha$ ($\alpha=1, \cdots, N$).
The electron spin is $s=\uparrow,\downarrow$ and the $c^\dagger_{k\alpha as} (c^{}_{k \alpha as})$ are the usual fermionic creation (annihilation) operators, with $\{ c^\dagger_{k\alpha as},c^{}_{k'\alpha' a's'}\} =\delta_{kk'}\delta_{\alpha \alpha'}\delta_{aa'}\delta_{ss'}$.
The threshold energy to open the transversal propagation mode $a$ in the lead $\alpha$ is $E_{\alpha as}$. We assume free motion in the direction along the leads. Hence, $E_{k\alpha as}=E_{\alpha as} + \hbar^2k^2/2m^*$, where $m^*$ is the electron effective mass.  The electrons at the lead $\alpha$ are in thermal equilibrium with the reservoir at temperature
$T$ to which the lead is connected. This reservoir is characterized by a chemical potential $\mu_\alpha$.

The conductor Hamiltonian reads \be \label{eq:Hdot} {\cal H}_{\rm
C}=\sum_{\mu \nu,s} \left[H_{\rm C}\right]_{\mu\nu}
d^{\dagger}_{\mu s}d^{}_{\nu s} , \ee where $d^\dagger_{\mu s}
(d_{\mu s})$ creates (annihilates) an electron at the $\mu$-th
state of an arbitrary basis $\{\nu\}$ that spans the conductor
eigenstates. Since we consider ${\cal H}_{\rm C}$ as a bilinear
operator, electron-electron interactions are only taken into
account in the mean-field level. This is a good approximation,
provided the system is open,\cite{Aleiner02} and hence neither
charging nor electronic correlations effects are expected to play
an important role.

The term that couples the leads to the conductor is
\be
\label{eq:Hcoupling}
{\cal H}_{\rm LC} = \sum_{k\alpha a,\mu,s} \left[ V^{}_{k \alpha a,\mu}
c^{\dagger}_{k \alpha a s} d^{}_{\mu s}  + \mbox{H.c.} \right]\;.
\ee

When there is a difference between the reservoirs' electro-chemical potentials
$\mu_\alpha$, the system is driven out of equilibrium and a current flows.
In the stationary regime a time-independent non-equilibrium self-consistent
electrostatic potential $U({\bf r})$ is formed. It depends on the applied bias,
as well as on the system geometry and material properties. In most of the paper,
we assume $U({\bf r})$ to be local. A non-local $U$ is discussed in Section \ref{sec:HF}.

The Hamiltonian ${\cal H}_{\rm L}$ of Eq.~(\ref{eq:Hlead}) assumes free
propagation in the leads. Alternatively, without significant increase in
complexity, it can also represent periodic semi-infinite leads.\cite{Hernandez07}
In any of these events, the surface separating conductor and leads has to
be chosen in such a way that, at any given lead $\alpha$, $U({\bf r}) \approx V_\alpha$.
As a consequence, the spatial dependence of $U({\bf r})$ is entirely accounted
for by ${\cal H}_{\rm C}$.
This construction not only limits the arbitrariness in defining the model Hilbert space,
but also guarantees a simple prescription for computing the characteristic potentials
$u_{\alpha\beta\cdots}({\bf r})$, as we shall discuss in Section \ref{sec:expansion}.
In the standard DFT approach for molecular electronics, although not emphasized, this
is the key notion behind defining an ``extended molecule" and it is essential to ensure
gauge invariance.\cite{Xue02}

There is a more basic principle behind the above construction \cite{Buttiker93} than just
simplifying calculations: The surface ${\cal S}$ defines the leads region at a position where no
electrical field lines penetrate its surface. Hence, the charge within the volume
${\cal V}$ enclosed by ${\cal S}$ is constant.

For the sake of simplicity we restrict our considerations to weak magnetic fields, or
more precisely, to systems where we can neglect spin-orbit and Zeeman interactions and
consider only orbital effects due to an external magnetic field. Hence, in what follows,
except for Sec.~\ref{sec:HF}, we omit the spin index and replace the sums over spin
projections by their degeneracy factors.

We also do not explicitly include the possibility of capacitive couplings in our model.
In pumping experiments, such kind of coupling is likely to dominate
the transport, \cite{pumpingexp} as discussed in Refs.~\onlinecite{pumpingtheo}.
In dc nonlinear transport, the effect of setting a fixed back gate voltage and, hence,
defining bias mode, can be relevant in nonlinear conductance of quantum
dots.\cite{Polianski07}
We stress that both the scattering and the NEGF approaches can easily accommodate
situations where the number of leads does not coincide with the number of gate voltages $\{V_\alpha\}$ and/or devices with more than one conductor.

\section{The NEGF approach}
\label{sec:I-V}

For elastic processes, the electronic current at the leads can be written in
terms of the conductor Green's functions, \cite{Meir92} namely
\be
\label{eq:I(Gr,G<)}
I_{\alpha} = -\frac{2e}{\hbar} \int^{\infty}_{-\infty} \frac{d E}{2 \pi}
\mbox{Im} \mbox{Tr} \Big\{ \mathbf{\Gamma}_{\alpha} \left[ \mathbf{G^{<}}(E) +
f_{\alpha} (E) \mathbf{G}^{r} (E) \right] \Big\},
\ee
where symbols in bold face correspond to matrices whose rows and columns are
states of the conductor basis set $\{\mu\}$. The Green's functions $G^r_{\mu\nu}(E)$
and $G^<_{\mu\nu}(E)$ are the Fourier transforms of $G^r_{\mu\nu}(t-t^\prime)=
-(i/\hbar) \theta(t-t^\prime)\langle \{d_\mu^{}(t),d_\nu^{\dagger}(t^\prime)\}\rangle$
and $G^<_{\mu\nu}(t- t^\prime) = (i/\hbar)\langle d_\nu^\dagger(t^\prime)
d_\mu^{}(t)\rangle$ respectively, where $\langle \cdots \rangle$ is defined as
standard.\cite{Haug96}
The decay width or line width matrix elements are given by
\be
\label{eq:decaywidth}
\left[\Gamma_\alpha\right]_{\mu\nu} = 2\pi \sum_{a\in \alpha}
V^{}_{\mu, k\alpha a} \,\rho^{}_{k\alpha a} (E)\,V^{*}_{\nu, k\alpha a}\;,
\ee
where $\rho_{\alpha a}(E)$ is the density of states of mode $a$ at the $\alpha$ contact.

Since the conductor Hamiltonian ${\cal H}_{\rm C}$ is a bilinear operator, the exact conductor Green's functions can be obtained in closed form using, for instance, the equations-of-motion method. \cite{Meir92} (For a recent review, see Ref.~\onlinecite{Hernandez07}.) In the energy representation, the Green's function $G^{r(a)}_{\mu\nu}$ is given by
\be \label{eq:grafinal}
{\bf G}^{r(a)}(E) = \left[ E {\bf I} - {\bf H}_{\rm C} - \mathbf{\Sigma}^{r(a)}(E)
\right]^{-1} \;.
\ee
The conductor lesser Green's function $G^{<}_{\mu\nu}$ follows directly from
the Dyson equation \cite{Haug96}
\be
\label{eq:Glesser}
{\bf G}^<(E) = {\bf G}^r(E)\mathbf{\Sigma}^<(E){\bf G}^a(E).
\ee
For the sake of definiteness the basis set is truncated and the matrices have
rank $M$. ${\bf I}$ is the identity matrix. (Later on we shall also use conductor
Green's functions in the coordinate representation and drop the boldface notation.)

The self-energy matrix elements read
\be \label{eq:selfenergy}
\Sigma^{}_{\mu \nu} (E) \equiv \sum_{k\alpha a} V^{}_{\mu, k\alpha a}\,g^{}_{k \alpha a}(E)
 \,V^{*}_{\nu, k \alpha a} \;,
\ee
where $\Sigma^{r(a)}$ (and $\Sigma^<$) are obtained by identifying the free
electron Green's function in the leads $g^{}_{k\alpha a}$, with
\be
\label{eq:free-g<}
g^<_{k\alpha a}(E)  = i f_\alpha(E)\delta(E - E_{k\alpha a})
\ee
and
\be
\label{eq:free-gra}
g^{r(a)}_{k\alpha a} (E) = \mp i \pi \delta(E - E_{k\alpha a}) + {\rm PV} \frac{1}{E - E_{k\alpha a}},
\ee
where PV stands for principal value integral.

The coupling matrix elements $V_{\mu,k \alpha a}$ are, in general, smooth
functions of the wave number $k$ and, hence, of $E_{k\alpha a}$.
Using (\ref{eq:free-gra}) and assuming that $\rho_{\alpha a}(E)$ has a broad band
width and a smooth energy dependence, the matrix elements $\Sigma^{r(a)}_{\mu\nu}$
become energy independent and read
\be
\label{eq:selfenergy_simple}
\mathbf{\Sigma}^{r(a)} \approx \pm\frac{i}{2} \mathbf{\Gamma}\equiv \pm \frac{i}{2} \sum_\alpha \left[\mathbf{\Gamma}_\alpha\right]\;. \ee
For situations where the broad and flat band approximation does not hold, the results we obtain for the $I-V$ characteristics have to be modified in a straightforward way, as indicated later on.

In analogy, the self-energy matrix elements $\Sigma^<_{\mu\nu}$ are given by
\ba
\label{eq26}
\Sigma^{<}_{\mu \nu} (E) & =& \sum_{\alpha=1}^N \sum_{ka}V^{}_{\mu, k\alpha a}\,g^{<}_{k\alpha a}(E)
\,V^{*}_{\nu, k\alpha a}
\nonumber\\
&\equiv&\sum_{\alpha=1}^N \left[\Sigma^{<}_\alpha(E)\right]_{\mu \nu}  \;.
\ea
Within the wide-band approximation, one arrives to
\be
\mathbf{\Sigma}_\alpha^<(E) \approx i f_\alpha(E)\mathbf{\Gamma}_\alpha\;.
\ee

Inserting ${\bf G}^r$ from Eq.~(\ref{eq:grafinal}) and ${\bf G}^<$ from
Eq.~(\ref{eq:Glesser}) into Eq.~(\ref{eq:I(Gr,G<)}), we write the current
$I_\alpha$ as in Eq.~(\ref{eq:LBscattering}),
\be
\label{eq:LB}
I_{\alpha}=-\frac{2 e}{h}\sum_{\beta=1}^N \int_{-\infty}^{\infty} \!
dE\, f_\beta(E)\, T_{\alpha\beta}(E,\{V_\gamma\}) \;.
\ee
with transmission coefficients given by
\be
\label{eq:AHernandez}
T_{\alpha\beta}(E,\{V_\gamma\}) = {\rm Tr}\Big[\mathbf{\Gamma}_\alpha
{\bf G}^r(E)(\mathbf{\Gamma} \delta_{\alpha\beta} - \mathbf{\Gamma}_\beta) {\bf G}^a(E)\Big]\;.
\ee
By means of the useful relation
\be
\label{eq:useful}
{\bf G}^a - {\bf G}^r = i{\bf G}^a\mathbf{\Gamma}{\bf G}^r =
-i{\bf G}^r\mathbf{\Gamma}{\bf G}^a,
\ee
we obtain $\sum_{\alpha=1}^N T_{\alpha\beta}=0$, and show that Eq.~(\ref{eq:LB})
satisfies current conservation, $\sum_{\alpha=1}^N I_\alpha= 0$.

As pointed out in Ref.~\onlinecite{Wang99}, the current $I_\alpha$ given by Eq.~(\ref{eq:LB})
is invariant under a global shift of the potential, that is, $V_\gamma \rightarrow V_\gamma + V_0$ for all $\gamma$'s and $U\rightarrow U+V_0$. This is not sufficient to prove that the formalism is gauge invariant. We still have to show that the same condition holds for the electron density $n({\bf r})$.
This is done in what follows.

The applied voltages $\{V_\gamma\}$ control the conductor charge distribution
$n({\bf r})$ and the electrostatic potential $U({\bf r})$. The latter, in turn,
enters the calculation of the conductor Green's functions.
Both quantities, $n({\bf r})$ and $G^<$, are related by
\be
n({\bf r}, t) = \sum_s\left\langle \psi_s^\dagger ({\bf r}, t) \psi_s({\bf r}, t)\right\rangle
= - 2 i \hbar \langle {\bf r} | G^< (t, t) | {\bf r} \rangle \,,
\ee
where the factor 2 account for the spin degeneracy.
By taking $\psi_s({\bf r}, t)$ inside the conductor as $\psi_s({\bf r}, t)
= \sum_{\mu} d_{\mu s}(t) \langle \mu| {\bf r}\rangle$, we obtain $n({\bf r}, t)$
in matrix representation, namely,
\be
\label{eq:density.vs.t}
n({\bf r}, t) = -2 i \hbar \sum_{\mu \nu} \langle {\bf r} | \mu \rangle G^<_{\mu\nu}(t,t)\langle \nu | {\bf r} \rangle.
\ee
For stationary processes, where $G^<(t,t) = G^<(0)$, the electronic density becomes
\be
\label{eq:density}
n({\bf r}) = - 2i \int_{-\infty}^{\infty} \frac{dE}{2\pi} \langle {\bf r}| G^<(E)|{\bf r} \rangle \;,
\ee
with obvious matrix representation. In order to close the calculational procedure, a relation between $n$ and $U$ is needed. This can be done at different approximation levels.

In the Hartree approximation, the electronic density $n({\bf r})$ and  $U({\bf r})$
are related by
\ba
\label{eq:Poisson}
\nabla^2 U({\bf r}) & = & -4\pi e\; n({\bf r})
\nonumber \\
&=& 8\pi i e \int_{-\infty}^{\infty} \frac{dE}{2\pi} \langle {\bf r} |G^<(E)| {\bf r} \rangle \;,
\ea
The boundary conditions are obtained by recalling that, by construction, $U({\bf r})$
is constant outside the conductor: $U({\bf r}) =V_\alpha$ when ${\bf r}$ is taken at the
lead $\alpha$.
In addition, the problem must to be solved self-consistently, namely, $U({\bf r})$ enters
the Hamiltonian ${\cal H}_{\rm C}$, which determines ${\bf G}^<(E)$, that in turn gives
$U{(\bf r})$.

Equation (\ref{eq:density}) leads to a local description of the electrostatic potential. In Section \ref{sec:HF}, we show how NEGF deals with a non-local potential due to the exchange interaction.

Equation (\ref{eq:density}) also plays a key role in the standard implementations of
the density functional theory (DFT) in molecular electronics (see, for instance,
Ref.~\onlinecite{Xue02} for a review). In DFT, $U({\bf r})=U[n({\bf r})]$ is considered as
a functional of $n({\bf r})$, containing exchange and correlation interactions in addition
to the Hartree one. Accordingly, the single-particle states $\{\mu\}$ become Kohn-Sham
orbital states.
Although this is a very appealing construction, it is not
as sound, from the conceptual point of view,
as the derivation presented here: DFT is not a mean-field theory and a bilinear
Hamiltonian, like Eq.~(\ref{eq:H}), is not one of its underpinning elements. A good
discussion about the shortcomings of the standard DFT approach to conductance
can be found in Ref.~\onlinecite{Koentopp07}.

We conclude this Section by stressing that in local approximation schemes Eq.~(\ref{eq:density})
is manifestly gauge invariant, provided the partition given by Eq.~(\ref{eq:H}) satisfies the
conditions discussed in Section \ref{sec:model}. In this case, any global voltage shift can be
absorbed by the energy integration, provided $U({\bf r})$ is calculated self-consistently.
For systems where electronic correlations are build across the partition ${\cal S}$, gauge
invariance calls for a more careful analysis. This is the case, for instance, in Kondo systems.
In the mean field limit, discussed here, such correlations are absent. In DFT-NEGF, any
semi-local functional of the exchange and correlation functional, $U_{xc}[n({\bf r})]$, allows
for a partition $\cal S$ and hence preserves gauge invariance, as nicely discussed in
Ref.~\onlinecite{Koentopp07}. Nonlocal interactions present in the exact XC functional
jeopardize the partition construction and spoil gauge invariance due to an XC contribution
to the characteristic potentials in the contacts.\cite{Stefanucci04,Koentopp07} In such
situations, a partition free approach, like the one discussed in Ref.~\onlinecite{Cini80}
is more suited.

\section{Linear and Nonlinear conductance coefficients}
\label{sec:expansion}

In this Section we present a systematic approach to calculate the conductance
coefficients ${\cal G}_{\alpha\beta\gamma\cdots}$ and discuss some of their
properties. For that purpose, we expand both $f_\alpha(E)$ and
$T_{\alpha \beta}(E,U({\bf r}))$, in Eq.~(\ref{eq:LB}), as powers of the voltages $\{ V_\alpha\}$.

We start writing the retarded (advanced) Green's function as
\be
G^{r(a)}(E) = \frac{1}{E - H_0 - eU - \Sigma^{r(a)}(E)}
\ee
without choosing a particular representation.
Next we expand $G^{r(a)}$ in terms of the differences between the non-equilibrium and
equilibrium $U$ and $\Sigma$. As a result, we obtain the Dyson equation
\be
\label{eq:DysonVeff}
G^{r(a)} = G^{r(a)}_0 + G^{r(a)}_0 V_{\rm eff} G^{r(a)} \;
\ee
where the effective perturbation potential $V_{\rm eff}$ is given by
\ba
\!\!\!\!\!\!\!\!
V_{\rm eff}(E) &=& eU - eU_{\rm eq} +
\nonumber\\\ && \!\!\!\!\!
\sum_\alpha\sum_{a \in \alpha} \Big[
\Sigma^{r(a)}_a(E - eV_\alpha) - \Sigma^{r(a)}_a(E)\Big],
\ea
and the equilibrium Green's function by
\be
G^{r(a)}_0(E) = \frac{1}{E - H_0 - eU_{\rm eq} - \Sigma^{r(a)}_{\rm eq}(E)}.
\ee

We now proceed by writing the self-energy
\be
\Sigma^{r(a)}(E) = \sum_\alpha
\sum_{a \in \alpha} \Sigma^{r(a)}_a(E - eV_\alpha)
\ee
as
\ba
\label{eq40}
\!\!\!\Sigma^{r(a)}_a (E-eV_\alpha)&=&
\Sigma^{r(a)}_a(E) - eV_\alpha \left.\frac{\partial \Sigma^{r(a)}_a}{\partial E}\right|_{V_\alpha = 0}
\nonumber\\&&
 +
e^2V_\alpha^2 \left.\frac{\partial^2 \Sigma^{r(a)}_a}{\partial E^2}\right|_{V_\alpha = 0} + \cdots
\ea
where $\Sigma^{r(a)}(E)=\Sigma^{r(a)}_{\rm eq}(E)$. From Eq. (\ref{eq40}) we see that,
in the flat band approximation, retarded and advanced self-energies do not depend on $\{V_\alpha\}$, since
\be
\left.\frac{\partial \Sigma^{r(a)}_a}{\partial E}\right|_{V_\alpha = 0} \approx
\left.\mp \frac{i}{2}\frac{\partial \Gamma_a}{\partial E}\right| = 0\;.
\ee
Hence, $V_{\rm eff}$ depends only on the characteristic potentials, namely
\ba
\label{eq:Veff-simple}
V_{\rm eff}= e \sum_\alpha u_\alpha V_\alpha + \frac{1}{2}e\sum_{\alpha\beta} u_{\alpha\beta} V_\alpha V_\beta +\cdots.
\ea
Inserting the above expression into (\ref{eq:DysonVeff}) we formally obtain $G^{r(a)}$
to arbitrary order in $V$.

We now turn our attention to the electronic density imbalance $\delta n({\bf r}) =
n({\bf r}) - n_{\rm eq}({\bf r})$, that ultimately allows us to calculate the characteristic
potentials $u_{\alpha\beta \cdots}$.
To obtain $\delta n({\bf r})$, we expand $G^<=G^r \Sigma^< G^a$ in powers of $\{V_\alpha\}$
taking, as above, the wide flat band limit. In this approximation, the lesser self-energy
$\Sigma^<$, Eq.~(\ref{eq26}), reads
\ba
\Sigma^<(E)
&=& i \sum_\alpha f_0(E - eV_\alpha) \Gamma_\alpha
\nonumber\\&=&
i \left( f_0\Gamma - e \frac{\partial f_0}{\partial E} \sum_\alpha V_\alpha  \Gamma_\alpha + \cdots \right) \;.
\ea
Finally, $G^<$ reads
\ba
\label{eq:G<expansion}
G^< &&\!\!\!\!\!\!= i f_0 G_0^r \Gamma G_0^a
\nonumber\\
&&\!\!\!\!\!\!
- e\sum_\alpha V_\alpha \!\left[ i \frac{\partial f_0}{\partial E} G_0^r \Gamma_\alpha G^a_0
+f_0 \left(G_0^r u_\alpha G_0^r - G_0^a u_\alpha G_0^a \right) \right]
\nonumber\\
&&\!\!\!\!\!\!
+ O(V^2) \;.
\ea
Close to equilibrium, when $\{V_\alpha\}\rightarrow 0$, $G^< \rightarrow G^<_0=-2i f_0 {\rm Im}G^r_0$, as given by the fluctuation-dissipation theorem. \cite{Haug96}
We use Eqs.~(\ref{eq:density}) and (\ref{eq:G<expansion}) to write the electronic density
as
\be
\label{eq:powersn}
n({\bf r}) = \sum_\ell n^{(\ell)}({\bf r})
\ee
where the $\ell$'s stand for the implicit powers of $V^\ell$. Note that $n^{(0)}({\bf r}) = n_{\rm eq}({\bf r})$.

We are now ready to identify the conductance coefficients ${\cal G}_{\alpha\beta\gamma \cdots}$
order by order: By plugging Eqs.~(\ref{eq:DysonVeff}) and (\ref{eq:Veff-simple}) into (\ref{eq:AHernandez})
we obtain the transmission $T_{\alpha\beta}$ in terms of equilibrium Green's functions and the characteristic
potentials. The later can be computed from $G^<$, as given by Eq.~(\ref{eq:G<expansion}), with the help, for instance, of the Hartree equation (\ref{eq:Poisson}).

\subsection{Linear conductance coefficients}

To linear order in $V$, the current at the contact $\alpha$ is
\be
I_\alpha^{(1)} = \sum_\beta {\cal G}_{\alpha\beta} V_\beta \;.
\ee
The conductance coefficients ${\cal G}_{\alpha \beta}$ are obtained from the linear
expansion of the current (\ref{eq:LB}) in the applied voltages $V_\alpha$.
They read
\be
\nonumber
{\cal G}_{\alpha \beta} = -\frac{2e^2}{h} \int_{-\infty}^\infty
dE\left(-\frac{\partial f_0}{\partial E}\right) T_{\alpha\beta}(E, \{V_\gamma\}=0)
\;.
\ee
where
\be
\label{eq:Landauer-Buttiker}
T_{\alpha\beta}(E, \{V_\gamma\}=0)={\rm Tr}\Big[\mathbf{\Gamma}_\alpha {\bf G}_0^r(E)
(\mathbf{\Gamma} \delta_{\alpha\beta} - \mathbf{\Gamma}_\beta) {\bf G}_0^a(E)\Big]\;.
\ee
which is identical to the multi-lead Landauer-B\"uttiker formula, \cite{Buttiker88} as
it can be verified following, for instance, the path presented in
Ref.~\onlinecite{FisherLee81}.

Owing to physical considerations the linear conductance coefficients ${\cal G}_{\alpha \beta}$
follow some simple sum rules. \cite{Buttiker93}
Current conservation implies that $\sum_\alpha {\cal G}_{\alpha\beta} = 0$. This sum rule
is automatically satisfied by Eq.~(\ref{eq:Landauer-Buttiker}), since Eq.~(\ref{eq:LB}) does
it to all orders in $V$. Current invariance under a global voltage shift $V_\alpha \rightarrow
V_\alpha + V_0$ leads to $\sum_\beta {\cal G}_{\alpha\beta} = 0$. This is fulfilled, since
$\sum_\beta T_{\alpha\beta}(E,\{V_\gamma\}) =0$.

Equations (\ref{eq:density}) and (\ref{eq:G<expansion}) give the electronic density as
\be
n_{\rm eq}({\bf r}) = -\frac{2}{\pi}\int_{-\infty}^{\infty}dE f_0(E){\rm Im}\langle {\bf r}| G^r_0(E)|{\bf r}\rangle \;.
\ee
It is worth remarking that, even in the Hartree approximation, depending on the system,
the computation of $U_{\rm eq}({\bf r})$ can already be a formidable computational task.
In mesoscopic physics, due to the chaotic and/or weakly disordered nature of the
addressed systems, quantitative results can be obtained by a statistical
treatment\cite{Aleiner02} using random matrix theory or diagrammatic techniques.
In molecular electronics a full electronic structure calculation is already
necessary.

An important symmetry of linear transport is unveiled by considering an external
magnetic field. The linear conductance coefficients fulfill the Onsager-Casimir
reciprocity relations under magnetic field inversion. \cite{Onsager}
This is indeed the case of Eq.~(\ref{eq:Landauer-Buttiker}).\cite{Buttiker86,Buttiker88}
Using ``microreversibility"
\be
\label{eq:microreversibility}
G^{r(a)}_0(-B) = [G^{r(a)}_0(B)]^T
\ee
and the cyclic properties of the trace in (\ref{eq:Landauer-Buttiker}), one can show that
\be
\label{eq:Onsager-Casimir}
{\cal G}_{\alpha\beta}(-B) = {\cal G}_{\beta\alpha}(B)
\ee
for $\alpha \neq \beta$. For the diagonal coefficients, where  $\alpha=\beta$,
one can either use current conservation and (\ref{eq:Onsager-Casimir}), or directly
use Eq.~(\ref{eq:useful}) to show that ${\cal G}_{\alpha \alpha}(B) =
{\cal G}_{\alpha \alpha}(-B)$.

In the two-terminal case, the conductance itself is an even function of the applied magnetic
field, namely,
\be
\label{eq:2terminalOnsager}
{\cal G}_{12}(-B) = {\cal G}_{12}(B),
\ee
which is a more stringent symmetry than the reciprocity relation for the general
multi-lead case.
In the scattering approach, Eq.~(\ref{eq:2terminalOnsager}) can be viewed as
a consequence of the $S$-matrix unitarity, \cite{Buttiker86,Buttiker88}
whereas using NEGF it follows from current conservation and from
${\cal G}_{\alpha \alpha}(B) = {\cal G}_{\alpha \alpha}(-B)$.
The even symmetry of ${\cal G}$ with respect to $B$ has been experimentally
established \cite{Benoit86,Yacoby} and has important implications for interference
experiments in two-terminal mesoscopic rings: It does not allow one to measure
phase differences, since in the observed Aharonov-Bohm conductance oscillations
the phase shift is locked either to 0 or to $\pi$. \cite{Yacoby,Yeyati95}

\subsection{Second order terms}
\label{sec:second_order}

The next order current term, in powers of $V$, is
\be
I^{(2)}_\alpha = \sum_{\beta\gamma} {\cal G}_{\alpha\beta\gamma}V_\beta V_\gamma \;.
\ee
The coefficient ${\cal G}_{\alpha\beta\gamma}$, obtained by expanding $\sum_\beta
f_\beta T_{\alpha\beta}$ of Eq.~(\ref{eq:LB}), is formally given by Eq.~(\ref{eq:Gabc}):
\begin{widetext}
\be
\label{eq:Gabc_implicit}
{\cal G}_{\alpha \beta \gamma} = -\frac{2e^3}{h} \int_{-\infty}^\infty
dE\left(-\frac{\partial f_0}{\partial E}\right)
\int_{\cal V} d{\bf r} \;
\left[u_{\gamma}({\bf r})-\frac{\delta_{\beta \gamma}}{2}\right]
\left.\frac{\delta T_{\alpha\beta}}{e\delta U({\bf r})}\right|_{\{V_\delta\}=0}\;.
\ee
Its explicit expression in terms of equilibrium Green's functions is \cite{Wang99}
\be
\label{eq:Gabc_explicit}
{\cal G}_{\alpha\beta\gamma} =
-\frac{2e^3}{h} \int_{-\infty}^{\infty}\! dE \left(- \frac{\partial f_0}{\partial E}\right)
{\rm Tr}\left\{\Gamma_\alpha G_0^r\left[\left(u_\gamma
- \frac{\delta_{\beta\gamma}}{2}\right)G_0^r (\Gamma\delta_{\alpha\beta} - \Gamma_\beta) +
(\Gamma\delta_{\alpha\beta} - \Gamma_\beta) G^a_0 \left(u_\gamma - \frac{\delta_{\beta\gamma}}{2}\right)
\right]G^a_0 \right\}.
\ee
\end{widetext}
Note that Ref.~\onlinecite{Wang99} accounts for an energy dependence in the self-energy and
presents a slightly more general equation for ${\cal G}_{\alpha\beta\gamma}$.

Current conservation implies that $\sum_\alpha {\cal G}_{\alpha\beta\gamma}=0$. It is straightforward
to verify that (\ref{eq:Gabc_explicit}) satisfies this sum rule.
The current invariance under a global shift of the applied voltages gives a second sum rule, \cite{Buttiker93} namely, $\sum_{\gamma}( {\cal G}_{\alpha\beta\gamma}+{\cal G}_{\alpha\gamma\beta})=0$. We show that the coefficients ${\cal G}_{\alpha\beta\gamma}$ of Eq.~(\ref{eq:Gabc_explicit}) also respect this sum rule, provided that $\sum_\gamma u_\gamma({\bf r}) = 1$. The latter is also be directly inferred from gauge invariance, see Section \ref{sec:scattering}.

We now turn our attention to the electron density $n^{(1)}({\bf r})$ of (\ref{eq:powersn}).
The expansion of $G^<$ in powers of $V$ and (\ref{eq:density}) give
\be
\label{eq:n1}
\!\!n^{(1)}({\bf r}) = e \sum_\alpha V_\alpha \left[\int_{\cal V} d{\bf r}^\prime \,
\Pi  ({\bf r},{\bf r}^\prime )u_\alpha({\bf r}^\prime) - \frac{dn({\bf r},\alpha)}{dE}\right],
\ee
where
\be
\Pi ({\bf r},{\bf r}^\prime) = - 2i \!\int_{-\infty}^{\infty}\! \frac{dE}{2\pi} f_0 \Big[
\langle {\bf r} | G^r_0 (E)|{\bf r}^\prime\rangle\langle{\bf r}^\prime|G^r_0 (E)|{\bf r}\rangle-
{\rm H.c.}
\Big]
\ee
is formally identified with the Lindhard function and
\be
\label{eq:NEGFinjectivity}
\frac{dn({\bf r},\alpha)}{dE} = 2 \int_{-\infty}^\infty \frac{dE}{2\pi}
\left(-\frac{\partial f_0}{\partial E}\right) \langle {\bf r} | G^r_0
\Gamma_\alpha G^a_0 | {\bf r} \rangle
\ee
is a partial local density of states, called injectivity in the scattering approach.
The compact form of (\ref{eq:NEGFinjectivity}) is due to the flat and wide band approximation. As discussed in Ref.~\onlinecite{Wang99}, $dn({\bf r},\alpha)/dE$ can be easily modified to account for a system specific energy dependence of self-energy $\Sigma$. Such corrections are potentially important in molecular electronics, where the details of the contacts should matter.

In the Hartree approximation, the characteristic potentials $u_{\gamma}({\bf r})$
are determined by
\be\label{eq:NEGFPoisson}
\!\!\!\nabla^2 u_\alpha({\bf r}) = 4\pi e^2
\left[\int_{\cal V} d{\bf r}^\prime \,
\Pi  ({\bf r},{\bf r}^\prime )u_\alpha({\bf r}^\prime) - \frac{dn({\bf r},\alpha)}{dE}\right]
\ee
with the boundary conditions discussed in Section \ref{sec:model}.
The above equation has the same structure as Eq.~(\ref{eq:scatPoisson}).
(We postpone a detailed comparison between both equations to the forthcoming Section.)
Both terms in the r.h.s. of Eq.~(\ref{eq:NEGFPoisson}) stem from of the electronic charge
imbalance $n^{(1)}({\bf r})$. \cite{Levinson89}

By using (\ref{eq:useful}) and integrating by parts, the important relation
\be
\label{SumInj=EmiT}
\sum_\alpha  \frac{dn({\bf r},\alpha)}{dE} = \int_{\cal V} \! d{\bf  r}^\prime\, \Pi({\bf r}, {\bf r}^\prime)
\ee
is obtained. This relation holds also beyond the flat and wide band approximation.
\cite{Wang99}

We now sum Eq.~(\ref{eq:NEGFPoisson}) over all leads $\alpha$ to write
\ba
\!\!\nabla^2 \sum_\alpha u_\alpha ({\bf r})= 4 \pi e^2\! \int_{\cal V} d{\bf r}^\prime \,
\Pi  ({\bf r},{\bf r}^\prime )\left[ \sum_\alpha u_\alpha({\bf r}^\prime) - 1 \right].
\ea
Recalling the boundary conditions for $u_\alpha$ we find that $\sum_\alpha u_\alpha = 1$, formally recovering one of the sum rules put forward by B\"uttiker \cite{Buttiker93}.
This is a quite simple way to prove that, in the Hartree approximation, the NEGF formalism is manifestly gauge invariant. Conversely, the scattering approach uses $\sum_\alpha u_\alpha = 1$ to obtain
(\ref{eq:scatPoisson}).

Another way to picture the sum rule $\sum_\alpha u_\alpha = 1$ is by observing that it automatically
guarantees that $n^{(1)}({\bf r})$ remains invariant under the global shift $V_\alpha \rightarrow V_\alpha + V_0$ (as shown to all orders in the previous section).

As experimentally established \cite{Lofgren04,Wei05,Marlow06,Zumbuhl06} and theoretically
discussed, \cite{Sanchez04,Spivak04} the Onsager-Casimir reciprocity relations do not hold for non-linear conductance.
In the formalism we present, this is manifest in
Eq.~(\ref{eq:Gabc_implicit}). While $\delta T_{\alpha\beta}/\delta U({\bf r})$
computed at $\{V_\gamma\}=0$ is even in magnetic field, in general $u_\gamma({\bf r},B)\neq u_\gamma({\bf r},-B)$. The later can be seen from Eq.~(\ref{eq:NEGFPoisson}):
Albeit the Lindhard function is even in $B$
\be
\Pi ({\bf r},{\bf r}^\prime; B)=\Pi ({\bf r},{\bf r}^\prime; -B),
\ee
as a consequence of the ``microreversibility" relation (\ref{eq:microreversibility})  $G^{r(a)}({\bf r},{\bf r}^\prime;E,B) = G^{r(a)}({\bf r}^\prime,{\bf r};E,-B)$,
in general
\be
\frac{dn({\bf r},\alpha,B)}{dE} \neq \frac{dn({\bf r},\alpha,-B)}{dE}.
\ee

\subsection{Arbitrary order}
\label{sec:arbitrary_order}

Recent experiments measured higher order conductance coefficients \cite{Leturcq06},
calling for a theoretical analysis of higher conductance coefficients. The general
expression for ${\cal G}_{\alpha\beta_1\cdots\beta_J}$ in terms of the ``generating
functional" $T_{\alpha\beta}$ is
\begin{widetext}
\ba
{\cal G}_{\alpha \beta_1 \cdots \beta_J} &=& -\frac{2e^2}{h}\int_{-\infty}^\infty dE \left(- \frac{\partial f_0}{\partial E}\right) \sum_{l=1}^{J} \sum_{n=0}^{J-l}\frac{(-1)^{l+1}}{l!} \delta_{\beta_1 \beta_2}\delta_{\beta_2 \beta_3}\cdots \delta_{\beta_{l-1} \beta_l}
 \nonumber \\ && \times
\int d{\bf r}_1 \cdots d{\bf r}_{l-1} \int d{\bf r}^{\prime}_1 \cdots d{\bf r}^{\prime}_n
\frac{\delta T_{\alpha \beta_1}}{\delta U({\bf r}_1)\cdots \delta U({\bf r}_{l-1})\delta U({\bf r}^{\prime}_1)\cdots \delta U({\bf r}^{\prime}_{n})}
K^{(n)}_{\beta_{l+1}\cdots \beta_J}({\bf r}^\prime_1, \cdots, {\bf r}^\prime_n),
\label{eq:GN}
\ea
where $K^{(n)}$ is defined as
\be
K^{(n)}_{\beta_{l+1}\cdots \beta_J}({\bf r}^\prime_1, \cdots, {\bf r}^\prime_J)=
\left.
\frac{1}{(J-l)!\,n!}\,\frac{\partial}{\partial V_{\beta_{l+1}}}\cdots\frac{\partial}{\partial V_{\beta_J}}
\Big[U({\bf r}^\prime_1) \cdots U({\bf r}^\prime_n)\Big]
\right|_{\{V_\gamma \}=0}
\ee
and related to the characteristic potentials by (\ref{eq:defu_char}). For $n=0$, we define
$K^{(0)}=\delta_{Jl}$.

Let us explain the structure of Eq.~(\ref{eq:GN}).
We identify the term containing $V$ to the power $J$ in the $I-V$ characteristics
Eq.~(\ref{eq:generalcurrent}) with the conductance coefficient
${\cal G}_{\alpha,\beta_1, \cdots \beta_J}$.
The $V^J$ term comes from the product of the expansions of $T_{\alpha\beta}$ and
$f_0$ in (\ref{eq:LB}).
The first sum in Eq.~(\ref{eq:GN}) run over $l$, the power of $V$ that stem from the
expansion of $f_0$.
The $J-l$ derivatives of $T_{\alpha\beta}$ with respect to $V$ give raise to higher order characteristic potentials and functional derivatives of the kind
$\delta T_{\alpha\beta}/\delta U({\bf r}^{\prime})$.
The second sum in Eq.~(\ref{eq:GN}) run over $n$, that represent the number of derivatives $\partial_{V_{\gamma}}$ that become $u_\gamma({\bf r}^\prime)\delta/\delta U({\bf r}^{\prime})$. The remaining $J-l-n$ derivatives are responsible for higher order characteristic potentials $u_{\beta_1\beta_2\cdots}$.
The expansion of $f_0$ is straightforward, but to factorize the $(-\partial_{E}f_0)$ term, we have to integrate by parts. As a result, we obtain additional $l-1$ functional derivatives acting on $T_{\alpha\beta}$.

Note that, as discussed in Sec. \ref{sec:scattering} the indices $\beta_1 \cdots \beta_l$ are not symmetrized. By combinatorial arguments it is not difficult to find the Taylor coefficients of the $I-V$ expansion.

Similar expressions have been obtained classically.\cite{Gaspard} The connection between the classical and quantum results is not clear yet.

The expression for the characteristic potential of order $J$ is obtained from the expansion of $G^<$ in Eq. (\ref{eq:Poisson})
\ba
\nabla^2 u_{\beta_1 \cdots \beta_J}({\bf r})&=&-8\pi e \int \frac{dE}{2\pi}
\frac{\partial}{\partial V_{\beta_1}}\cdots \frac{\partial}{\partial V_{\beta_J}} \sum_{k=0}^{J}
\langle {\bf r}| G_0^{r} \overbrace{V_{\rm eff} G_0^r  \cdots V_{\rm eff}G_0^r}
^{k \,{\rm terms}}\;
\sum_{\alpha}f(E-eV_{\alpha}) \Gamma_{\alpha}  \nonumber \\ &&
\left.  \times \sum_{j=0}^{J-k} G_0^{a}
\overbrace{V_{\rm eff}G_0^a \cdots V_{\rm eff}G_0^a}
^{j \,{\rm terms}} | {\bf r}\rangle \right|_{\{V_\gamma \}=0}
.
\label{eq:UN}
\ea
The effective potential $V_{\rm eff}=eU - eU_{\rm eq}$ does not change upon a global shift $V_\alpha \rightarrow V_\alpha + V_0$.

Equations (\ref{eq:GN}) and (\ref{eq:UN}) allow us to solve the non-linear problem to arbitrary order. We checked that, in lowest order, these equations lead to the results discussed in the previous sections, as they should. For the sake of illustration we explicitly show the third order conductance coefficient, namely,
\ba
{\cal G}_{\alpha \beta \gamma \delta} = -\frac{e^2}{h}\int_{-\infty}^\infty dE
&&\!\!\!\!\!\!\left(- \frac{\partial f_0}{\partial E}\right)
\left\{
\int d{\bf r}_1 \frac{\delta T_{\alpha \beta}}{\delta U({\bf r}_1)} u_{\delta \gamma}({\bf r}_1) \right.
\nonumber \\
+&&\!\!\!\!\!\!\int d{\bf r}_1 \int d{\bf r}_2 \left.\frac{\delta T_{\alpha \beta}}{\delta U({\bf r}_1)\delta U({\bf r}_2)}
\left[u_{\gamma}({\bf r}_1)u_{\delta}({\bf r}_2)-\delta_{\beta \gamma} u_{\delta}({\bf r}_1)
+ \frac{1}{3}\delta_{\beta \gamma}\delta_{\gamma \delta}\right]\right\},
\ea
and the corresponding equation for $u_{\alpha\beta}$
\be
\label{eq:uab}
\nabla^2 u_{\alpha\beta}({\bf r})=-4\pi e^2 \left[ -\frac{dn({\bf r})}{dE} u_{\alpha\beta}({\bf r}) + e \frac{d^2n({\bf r})}{dE^2}u_{\alpha}({\bf r})u_{\beta}({\bf r}) + e\frac{d^2n({\bf r},\beta)}{dE^2} \delta_{\alpha\beta}
- e\frac{d^2n({\bf r},\beta)}{dE^2} u_{\alpha}({\bf r}) - e\frac{d^2n({\bf r},\alpha)}{dE^2} u_{\beta}({\bf r}) \right].
\ee
For simplicity, $u_{\alpha\beta}$ is written in the Thomas-Fermi approximation. It is simple to explicitly show that $\sum_{\alpha}u_{\alpha\beta}=0$, respecting gauge invariance.

\end{widetext}
\section{Connection with the scattering-matrix approach}
\label{sec:connectionButtiker}

We now establish the equivalence between the NEGF results for nonlinear elastic electronic
transport and those obtained by the scattering approach, as formulated by B\"uttiker and
collaborators,\cite{Buttiker93} and summarized in Section \ref{sec:scattering}.

The underpinning elements of the scattering approach are:
(a) The scattering matrix is viewed as a function of the electron energy and a functional
of the non-equilibrium electrostatic potential in the conductor. $S(E,U({\bf r}))$ is
used to construct a generating functional to obtain the nonlinear conductance coefficients.
(b) Physical arguments are used to write a Poisson equation relating $U({\bf r})$ with
source terms expressed as functions of $S$. Both $S$ and $U$ are solved self-consistently.

We begin by writing the standard (equilibrium) resonance scattering matrix \cite{MahauxWeidenmueller}
\be
\label{eq:SHeidelberg}
S_{ba}(E) = \delta_{ba} - 2\pi i \sum_{\mu\nu}(\rho_a \rho_b)^{1/2} V^*_{\mu  b} [G^r_0(E)]_{\mu\nu} V_{\nu a}\;,
\ee
where $a$ and $b$ label propagating modes in the leads and the conductor Green's function is calculated for the equilibrium electrostatic potential $U_{\rm eq}({\bf r})$.
As the bias is increased $U({\bf r})$ is modified, as well as the conductor resonances and the channels thresholds. Provided that the applied bias does not add new physical processes to the transport problem, like for instance phonons, the ``equilibrium" $S$-matrix  can be generalized to
\be
\label{eq:Sgeneral}
S_{ba}(E,U({\bf r})) = \delta_{ba} - 2\pi i \sum_{\mu\nu}(\rho_a \rho_b)^{1/2} V^*_{\mu  b}
[G^r(E)]_{\mu\nu} V_{\nu a}\;.
\ee
Here the non-equilibrium $U({\bf r})$ is contained in the conductor Green's function $G^{r}$.
By recalling the definition of the decay widths $\Gamma$ , we obtain \cite{tracenote}
\be
A_{\alpha\beta}={\rm Tr}\left[{\bf 1}_{\alpha}\delta_{\alpha\beta}-{\bf S}^{\dagger}_{\alpha\beta}
{\bf S}^{}_{\alpha\beta}\right]
= -T_{\alpha\beta}
\ee
and conclude that the scattering and the NEGF approaches formally give identical
expressions for the current.

We now examine how $U({\bf r})$ is treated in both approaches, namely, we compare
Eqs.~(\ref{eq:scatPoisson}) and (\ref{eq:NEGFPoisson}).
We begin by comparing the injectivities. By making explicit the sums over channels,
the scattering approach injectivity (\ref{eq:injec}) becomes
\ba
\label{eq:injec0}
\frac{dn^{\rm s}({\bf r},\alpha )}{dE}
&=&-\frac{1}{2\pi i}\sum_{\beta}\int_{-\infty}^{\infty}\frac{dE}{2\pi}\left(- \frac{\partial f_0}{\partial E} \right)
\nonumber\\&\times &
\sum_{{a\in \alpha} \atop{b \in \beta}} \left[ S^{\dagger}_{ba}\frac{\delta S_{ba}}{e\delta U({\bf r})}-\frac{\delta S^{\dagger}_{ba}}{e\delta U({\bf r})}S_{ba}\right],
\ea
evaluated at $\{V_\gamma\}=0$. Equation (\ref{eq:Sgeneral}) renders a quite amenable path to calculate the functional derivative $\delta S^{\dagger}_{ba}/\delta U({\bf r})$: First, we
use $G^r \delta [(G^r)^{-1}] + (\delta G^r)(G^r)^{-1} = 0$ to write
\ba
\label{eq:derivfuncS}
\!\!\!\!\!\!\!\!\!\!\!\!\frac{\delta S_{ba}}{e \delta U({\bf r})} & = &
 +2\pi i (\rho_a\rho_b)^{1/2}
\nonumber\\
&&\!\!\!\!\!\!\!\!\!\!\!\!\!\!\!\!\!\!\!\!\!\!\!\!\!\!\!\!\!\!
\sum_{\mu\mu^\prime \nu \nu^\prime}V^*_{\mu b} [G^r(E)]_{\mu\mu^\prime} \frac {\delta [G^r(E)^{-1}]_{\mu^\prime
\nu^\prime}}{e \delta U({\bf r})}[G^r(E)]_{\nu^\prime \nu}V_{\nu a}.
\ea
Second, noting that $(G^r)^{-1} = E - H_0- eU + i\Gamma/2$ and $U_{\mu\nu} = \int d{\bf r}^\prime \langle \mu|{\bf r}^\prime \rangle U({\bf r}^\prime) \langle {\bf r}^\prime | \nu\rangle$, we readily write
\be
\label{eq:derivfuncG}
\frac {\delta }{e \delta U({\bf r})}[G^r(E)^{-1}]_{\mu\nu}= - \langle \mu|{\bf r} \rangle  \langle {\bf r} | \nu\rangle\;.
\ee
Plugging Eqs.~(\ref{eq:SHeidelberg}) and (\ref{eq:derivfuncS}) into (\ref{eq:injec0})
and using  (\ref{eq:derivfuncG}) we show that $d n^{\rm s}({\bf r},\alpha )/dE$ exactly
coincides with Eq.~(\ref{eq:NEGFinjectivity}), obtained within the wide and flat band approximation.

Following the same steps, we also find that the scattering approach emissivity is given by
\be
\frac{dn(\alpha, {\bf r})}{dE} = 2 \int_{-\infty}^{\infty}\frac{dE}{2\pi}\left(- \frac{\partial f_0}{\partial E} \right)
\langle {\bf r} | G_0^a \Gamma_\alpha G_0^r | {\bf r} \rangle.
\ee

The Thomas-Fermi approximation is key to have a closed calculational scheme in terms of scattering matrix. As standard, it is assumed that $u_{\alpha}({\bf r})$ shows a slower coordinate dependence than $\Pi ({\bf r},{\bf r}^{\prime})$ to write
\ba
\int_{\cal V} d{\bf r}^{\prime}\,\Pi ({\bf r},{\bf r}^{\prime})u_{\alpha}({\bf r}^{\prime})&\approx&
u_{\alpha}({\bf r})\int_{\cal V}  d{\bf r}^{\prime}\,\Pi ({\bf r},{\bf r}^{\prime})\nonumber\\
&\approx& u_{\alpha}({\bf r})\frac{dn({\bf r})}{dE}
\ea
where we use (\ref{SumInj=EmiT}). In this limit the NEGF Hartree equation (\ref{eq:NEGFPoisson}) reduces to the scattering Poisson equation (\ref{eq:scatPoisson}). It is interesting to note that the diagonal part of $\Pi ({\bf r},{\bf r}^{\prime})$ can be
written as
\ba
\Pi ({\bf r},{\bf r})
=-2i \!\int_{-\infty}^\infty \frac{dE}{2\pi} \left(- \frac{\partial f_0}{\partial E} \right) \Big[
 \langle {\bf r} | G^a_0(E)  - G^r_0 (E )|{\bf r}\rangle \Big]
\nonumber
\ea
which, integrating by parts and using (\ref{eq:useful}), gives
\be
\Pi ({\bf r},{\bf r}) =  \sum_\alpha \frac{dn(\alpha, {\bf r})}{d E}.
\ee
Note that Eq.~(\ref{eq:uab}) coincides with the corresponding one obtained in the scattering approach within the Thomas-Fermi approximation.\cite{Ma98}
This suggests that, within the approximations discussed in this  Section, both
approaches are equivalent to all orders.

\section{NEGF implementation for the Hartree-Fock approximation.}
\label{sec:HF}

In order to improve our mean field description of non-linear
conductance, we now include the exchange interaction term,
and consider the many-body problem in the Hartree-Fock approximation. The non-local
nature of the exchange interaction prevents the
description of the nonlinear conductance in terms of (local) characteristic
potentials defined in Eq.~(\ref{eq3}), making a description in terms of the scattering matrix approach unpractical. However, as before, it is possible to construct a self-consistent
expansion for the $I-V$ characteristics, and treat the problem by NEGF.

Let us consider the interacting Hamiltonian
\be
{\cal H}_{\rm C}= {\cal H}_{\rm 0}+ {\cal H}_{\rm int}
\ee
where
\be
{\cal H}_{\rm 0}=\sum_{\mu \nu,s} \left[H_{\rm 0}\right]_{\mu\nu}
d^{\dagger}_{\mu s}d^{}_{\nu s} ,
\ee
describes the single-particle terms of conductor Hamiltonian, and replaces the operator ${\cal H}_{\rm C}$ addressed so far, and
\be
{\cal H}_{\rm int}= \frac{1}{2}\sum_{ss^{\prime}}\sum_{\mu\nu\gamma\delta}V_{\mu\nu\gamma\delta}\;\;d^{\dagger}_{\mu s}
 d^{\dagger}_{\nu s^{\prime}} d_{\gamma s^{\prime}} d_{\delta s} \;,  \ee
with
\be
\label{79}
V_{\mu\nu\gamma\delta}=\int d{\bf r}_1 d{\bf r}_2 \phi^*_{\mu}({\bf r}_1)\phi^*_{\nu}({\bf r}_2)V({\bf r}_1-{\bf r}_2)\phi_{\gamma}({\bf r}_2)\phi_{\delta}({\bf r}_1)\; .
\ee

In the Hartree-Fock approximation, the interaction term reads \cite{Bruus-Flensberg2004}
\begin{widetext}
\be
\label{80}
H_{\rm int}^{HF}= \sum_{{\mu\nu\gamma\delta}\atop{ss^{\prime}}}V_{\mu\nu\gamma\delta}
\left[ \;
\big\langle d^{\dagger}_{\mu s} d_{\delta s} \big\rangle d^{\dagger}_{\nu s^{\prime}} d_{\gamma s^{\prime}}- \big\langle d^{\dagger}_{\mu s}d_{\gamma s^{\prime}}\big\rangle d^{\dagger}_{\nu s^{\prime}}  d_{\delta s}
+ \frac{1}{2} \big\langle d^{\dagger}_{\mu s}d_{\delta s}\big\rangle \big\langle d^{\dagger}_{\nu s^{\prime}} d_{\gamma s^{\prime}} \big\rangle
- \frac{1}{2} \big\langle d^{\dagger}_{\mu s} d_{\gamma s^{\prime}} \big\rangle \big\langle d^{\dagger}_{\nu s^{\prime}}  d_{\delta s} \big\rangle \right].
\ee
Since the $H_{\rm int}^{HF}$ is a bilinear operator, it is straightforward to obtain
 \be
 \label{81}
 \tilde{{\bf G}}^{r(a)}(E)=\left[({\bf G}^{r(a)}(E))^{-1}-{\bf H}_{\rm Int}^{HF}\right]^{-1}\; ,
 \ee
where the $H_{\rm int}^{HF}$ matrix elements are
\be
\left[H_{\rm int}^{HF}\right]_{\nu s , \gamma s^{\prime}}=\delta_{s,s^{\prime}}\sum_{\mu \delta}
\left[ V_{\mu\nu\gamma\delta}\sum_{s^{\prime\prime}} \big< d^{\dagger}_{\mu s^{\prime\prime}}d_{\delta  s^{\prime\prime}}\big> -
H_{\mu\nu\delta\gamma} \big< d^{\dagger}_{\mu s}d_{\delta s}\big>  \right]\; .
\ee
\end{widetext}
The lesser Green's function is given by
\be
\label{83}
\tilde{{\bf G}}^{<}(E)=\tilde{{\bf G}}^{r}(E) {\bf \Sigma}^{<}(E)  \tilde{{\bf G}}^{a}(E)
\ee
and the self consistent equations reads
\be
\label{84}
\big< d^{\dagger}_{\mu s}d_{\nu s^{\prime}}\big>=-i\int \frac{dE}{2\pi}\tilde{G}^{<}_{\nu s^{\prime}\; \mu s}(E)=-i \delta_{s,s^{\prime}}\int \frac{dE}{2\pi} \tilde{G}^{<}_{\nu \mu}(E).
\ee
Gauge invariance is shown to hold by the same arguments used in Section \ref{sec:I-V}.

Eqs. (\ref{81}), (\ref{83}) and (\ref{84}) provide the elements to write a power expansion of $I_{\alpha}$, in analogy to Section \ref{sec:expansion}. The non-local nature of the exchange interaction is encoded in Eqs. (\ref{79}) and (\ref{80}), and leads to a more involved self-consist scheme than that of Section \ref{sec:I-V}.
\section{Conclusions}
\label{sec:conclusions}

We studied the nonlinear phase coherent quantum electronic transport
properties of nanoscopic devices using the Nonequilibrium Green's function
method.
This method allows us to express the $I-V$ characteristics of a given
system to arbitrary powers of the applied voltages in terms of equilibrium
Green's functions.
We show that the formalism is gauge invariant, provided that $U({\bf r})$
is calculated self-consistently and the induced charge is well localized. The latter condition is key to partition the system as in Eq.~(\ref{eq:H}), the starting point of our discussion.

We explicitly establish a connection between the NEGF method and the scattering approach.
This is done by analyzing the first nonlinear contributions to the current,
namely, $I^{(2)}_\alpha$. We show that $I^{(2)}_\alpha$ obtained by NEGF
at the Hartree level reduces to the scattering matrix result in the Thomas-Fermi
limit (and by using the wide band approximation).
It should be noted that while in the scattering approach gauge invariance is used to
construct the Poisson equation, NEGF renders gauge invariance automatically.
These observations suggest that NEGF provides a framework for treating the many-body
problem at a more accurate level of approximation. In Sec.~\ref{sec:HF}, we discuss the Hartree-Fock approximation, very amenable to treat with NEGF, but clearly unsuited to the scattering approach, which is restricted to local potentials.

We also analyze the electronic transport symmetry with respect to magnetic field inversion. In particular, we discuss the consequences of ``microreversibility" for the conductance coefficients
${\cal G}_{\alpha\beta\cdots}$ using the NEGF method for the second order coefficients.
The general conclusions are the same as the ones obtained from the scattering approach.
We then generalize to the theory arbitrary order in $V$, which is useful to address nonlinear transport experiments, such as Ref.~\onlinecite{Leturcq06}.

In general, as the bias is increased, very quickly inelastic channels are opened. The inclusion of inelastic processes in the formalism and the development of approximation schemes to solve such problem is one of the next main goals to pursue.

\acknowledgments
We thank A. Wasserman for useful  discussions.
This work was supported by CNPq (Brazil), CAPES(Brazil), FAPERJ (Brazil), and the
Harvard's Institute of Quantum Science and Engineering.



\begin{thebibliography}{99}
\bibitem{tracenote}
We adopt the convention that a trace in an expression involving $S$-matrices refers to sums over
channels $a \in \alpha$ and $b \in \beta$, whereas traces over Green's functions are taken over
the conductor states $\{\mu\}$.
\end{thebibliography}
\end{document}